\documentclass[aps, pra, superscriptaddress, twocolumn, amsfonts, amsmath, amssymb, floatfix]{revtex4-2}
\usepackage{graphicx}
\usepackage{float}
\usepackage{sidecap}
\usepackage{dcolumn}
\usepackage{bm}
\usepackage{epsfig}
\usepackage{braket}
\usepackage{color}
\usepackage{ulem}
\usepackage{amsmath}
\usepackage{wasysym}
\usepackage[colorlinks=true, letterpaper=true, pdfstartview=FitV, linkcolor=blue, citecolor=blue, urlcolor=blue]{hyperref}
\usepackage{lipsum}

\begin{document}

\title{Bose-Einstein Condensates in a Synthetic Magnetic Field with Tunable Orientation}

\author{Fengtao Pang}
%\author{Fengtao Pang~\orcidlink{0000-0002-2784-0064}}
\thanks{These authors contributed equally to this work.}
\affiliation{Institute for Quantum Science and Technology, Department of Physics, Shanghai University, Shanghai 200444, China}

\author{Huaxin He}
%\author{Huaxin He~\orcidlink{0000-0003-4067-191X}}
\thanks{These authors contributed equally to this work.}
\affiliation{Institute for Quantum Science and Technology, Department of Physics, Shanghai University, Shanghai 200444, China}
\affiliation{College of Information Science and Engineering, Jiaxing University, Jiaxing, 314001, China}

\author{Yongping Zhang}
%\author{Yongping Zhang~\orcidlink{0000-0001-5090-9173}}
\email{yongping11@t.shu.edu.cn}
\affiliation{Institute for Quantum Science and Technology, Department of Physics, Shanghai University, Shanghai 200444, China}

\author{Chunlei Qu}
%\author{Chunlei Qu~\orcidlink{0000-0002-3080-8698}}
\email{cqu5@stevens.edu}
\affiliation{Department of Physics, Stevens Institute of Technology, Hoboken, NJ 07030, USA}
\affiliation{Center for Quantum Science and Engineering, Stevens Institute of Technology, Hoboken, NJ 07030, USA}

\begin{abstract}
We systematically investigate the ground state and dynamics of spinor Bose-Einstein condensates subject to a position-dependent detuning. This detuning induces three related quantities—a synthetic magnetic field, an angular velocity, and an angular momentum—which, due to trap anisotropy, may point in different directions. When the dipole frequencies along the three symmetric axes of the harmonic trap are degenerate, the dipole motion can decompose into two coupled transverse modes in the plane perpendicular to the synthetic magnetic field, and another decoupled longitudinal mode, enabling controllable Foucault-like precession or bi-conical trajectories depending on the excitation protocol. Furthermore, quenching the orientation of the synthetic magnetic field excites multiple coupled quadrupole modes. We develop a hydrodynamic theory whose predictions match well with Gross-Pitaevskii simulations. This study contributes to a deeper understanding of the effects of the synthetic magnetic field and the excitations of the collective mode in quantum fluids, providing a foundation for future developments in quantum simulation and high-precision sensing technologies.

\end{abstract}

\maketitle

\section{Introduction}\label{sec1}
% \textcolor{black}{You can start drafting it. For the first two paragraphs, I would consider writing a similar introduction part as the previous paper. For the last one or two paragraphs, we can highlight the novel aspect of this work, the motivation, and the main findings.}

Ultracold atomic gases offer an ideal platform for quantum simulation thanks to their high degree of controllability and versatility, enabling the exploration of exotic quantum phases and dynamics~\cite{anderson1995observation,o2002observation,lewenstein2007ultracold,RevModPhys.80.885}. By introducing specific interactions between light and neutral atoms, it becomes possible to realize synthetic magnetic fields and synthetic spin-orbit coupling~\cite{PhysRevLett.102.130401,lin2009synthetic,lin2011synthetic,lin2011spin,goldman2014light,huang2016experimental,wu2016realization}. Both magnetic fields and spin-orbit coupling are fundamental ingredients in topological physics~\cite{RevModPhys.82.3045}, which studies exotic quantum states that are robust under small perturbations. Leveraging these synthetic effects, recent experiments with ultracold atoms have successfully implemented topological models such as the Hofstadter-Harper model~\cite{PhysRevLett.111.185301,PhysRevLett.111.185302}, the Haldane model~\cite{jotzu2014experimental} and the Su-Schrieffer-Heeger model~\cite{leder2016real}, enabling investigations of their topological properties in controllable atomic systems~\cite{atala2013direct,lohse2016thouless,RevModPhys.91.015005,PhysRevLett.112.086401}.

The pioneering experimental realization of Raman-induced spin-orbit coupled Bose gases~\cite{lin2011spin}, featuring equal Rashba and Dresselhaus couplings, has attracted extensive experimental and theoretical interest~\cite{PhysRevLett.105.160403,PhysRevLett.107.150403,PhysRevA.84.063604,PhysRevLett.109.115301,PhysRevLett.108.010402,PhysRevLett.109.095301,PhysRevLett.108.225301,PhysRevA.85.053607,PhysRevA.86.063621,PhysRevA.86.041604,PhysRevA.87.053619,PhysRevA.87.031604,PhysRevA.87.063610,PhysRevLett.110.140407,PhysRevA.89.023629,PhysRevA.89.061605,PhysRevA.90.013616,PhysRevA.90.011602,PhysRevA.90.053606,PhysRevA.95.043623,qu2017spin1,li2017stripe,PhysRevA.95.033603,PhysRevA.95.043605,PhysRevA.98.013615,Hou_2018,li2019spin,PhysRevLett.125.260603,PhysRevResearch.2.033152,PhysRevA.106.023302,PhysRevLett.109.115301,tao2025imaginary}. Depending on the Raman coupling, the system's single-particle dispersion may exhibit a double-well structure or a single-well structure, corresponding to the plane-wave phase or the zero-momentum phase, respectively. Unlike conventional Bose-Einstein condensates (BECs), the presence of spin-orbit coupling breaks Galilean invariance~\cite{zhu2012exotic,zheng2013properties,PhysRevLett.114.070401}, leading to fundamentally altered superfluid characteristics. One notable example is the vanishing of the superfluid density near the transition from the plane-wave phase to the zero-momentum phase~\cite{PhysRevA.94.033635}. Another is the breakdown of irrotationality in the velocity field, which leads to many interesting collective dynamics~\cite{PhysRevLett.118.145302}.

In spin-orbit coupled Bose-Einstein condensates (SOC BECs), introducing a position-dependent detuning with a gradient perpendicular to the direction of the spin-orbit coupling can induce a synthetic magnetic field, which allows neutral atoms to experience an effective Lorentz force similar to that acting on charged particles in a real magnetic field~\cite{PhysRevLett.118.145302,PhysRevLett.120.183202}. This system, experimentally realized by the observation of quantized vortices at large detuning gradients~\cite{lin2009synthetic}, exhibits markedly different physics when the gradient remains below the vortex-formation threshold. In this regime, the condensate develops a rigid-body-like velocity field that drives nontrivial dynamics, including coupled dipole modes featuring Foucault-like precessional motion of the center-of-mass~\cite{PhysRevLett.120.183202}, as well as coupled quadrupole modes that reveal Hall effect~\cite{PhysRevA.108.053316,doi:10.1073/pnas.1202579109}, quantum gyroscopic effect~\cite{PhysRevA.108.053316}, Lissajous-like trajectories~\cite{PhysRevResearch.7.013219}, and intriguing spin-dependent expansion dynamics~\cite{PhysRevA.110.043307}. However, these investigations predominantly assume a fixed orientation of the synthetic magnetic field. Introducing a tunable field orientation represents a previously unexplored degree of freedom, which may uncover novel orientation-dependent phenomena and enrich the understanding of spin-orbit coupled systems.

In this work, we employ both spinor hydrodynamic (HD) theory and Gross-Pitaevskii (GP) simulations to systematically explore the ground-state structure and dynamical responses of SOC BEC in a synthetic magnetic field with tunable orientation. Our ground-state analysis reveals that trap anisotropy significantly modifies the rigid-body-like rotation: the angular velocity and angular momentum deviate from the orientation of the synthetic magnetic field. When the dipole frequencies along the three symmetric axes of the harmonic trap are degenerate, we find that the center-of-mass motion can be decomposed into distinct dynamics, leading to Foucault-like precession or its three-dimensional bi-conical extension depending on the excitation protocol. Furthermore, sudden quenches in the orientation of the synthetic magnetic field excite multiple quadrupole modes simultaneously, giving rise to rich dynamical phenomena.

The structure of this paper is as follows. Section~\ref{sec2} introduces the theoretical model for a SOC BEC subject to a synthetic magnetic field with tunable orientation, and outlines the derivation of the corresponding spinor HD equations. In Sec.~\ref{sec3}, we analyze the equilibrium properties of the condensate based on HD theory. In Sec.~\ref{sec4}, we explore the system's nonequilibrium dynamics by examining dipole and quadrupole modes following various quench protocols. Section~\ref{sec5} concludes the paper with a summary and discussion of the main findings.

\section{Model}\label{sec2}

We consider a two-component BEC confined in a 3D harmonic trap. The two internal states are coupled through a two-photon Raman transition, creating an effective spin-orbit coupling along the $x$-direction. The single-particle Hamiltonian of the system can be written as 
\begin{equation}
    \label{Hsingle}
    \begin{aligned}
            H_0=&\frac{(\hat{\bm{p}}-\hbar k_{0}\sigma_{z}\hat{\bm{e}}_x)^{2}}{2m}-\frac{\Omega}{2}\sigma_{x}+V_{\text{trap}} -\delta(y,z,\theta)\sigma_{z},
    \end{aligned}
\end{equation}
where $\hat{\bm{p}} = -i\hbar\bm{\nabla}$ is the momentum operator, $\hbar$ is the reduced Planck constant, and $k_0$ represents the recoil momentum imparted by the Raman laser. The corresponding recoil energy is $E_{\text{r}} \equiv {\hbar^{2}k_{0}^{2}}/{2m}$, where $m$ is the atomic mass. The system is confined in a 3D harmonic potential $V_{\text{trap}}=m(\omega_{x}^{2}x^{2}+\omega_{y}^{2}y^{2}+\omega_{z}^{2}z^{2})/2$ with trapping frequencies $\omega_{x,y,z}$. The position-dependent detuning is given by 
\begin{equation}
    \delta(y,z,\theta)=\eta k_{0}(y\sin\theta + z\cos\theta),
\end{equation}
where $\eta$ represents the magnitude of the detuning gradient
\begin{equation}
    \label{Gdetuning}
    \bm{\nabla}\delta = \eta k_0(0,\sin\theta,\cos\theta),
\end{equation}
with the orientation controlled by the parameter $\theta$. In general, for the zero-momentum phase, when a position-dependent detuning $\delta(\mathbf{r})$ is present, the lower energy band along the spin-orbit coupling direction can be approximated as~\cite{PhysRevA.79.063613,lin2009synthetic}:
\begin{equation}
    \label{dispersion}
    E(k_x)\approx\frac{\hbar^{2}}{2m^{*}} \left (k_x-\frac{2\delta(\mathbf{r}) }{\Omega-\Omega_{c}}k_{0}\right ) ^{2},
\end{equation}
where $\Omega_c = 4E_{\text{r}}$ represents the critical value of Raman coupling strength for the phase transition from plane-wave phase to zero-momentum phase. The effective mass of the zero-momentum phase is given by $m^{*} = m\left(1 - \Omega_c/\Omega\right)^{-1}$, quantifying the modification of the inertial response along the $x$-direction by the spin-orbit coupling~\cite{lin2011synthetic,PhysRevLett.109.115301}. Comparing Eq.~\eqref{dispersion} to the Hamiltonian of a charged particle moving in a magnetic field, one can introduce a synthetic vector potential $\bm{A}^* = A_x^*(\mathbf{r}) \hat{\bm{e}}_x$, where $A_x^*(\mathbf{r}) = \dfrac{2\delta(\mathbf{r})}{\Omega - \Omega_{c}}k_{0}$. The synthetic magnetic field, defined as the curl of such a synthetic vector potential, is given by
\begin{equation}  
    \bm{B}_{\text{syn}} = \hat{\bm{e}}_y\partial_z A_x^*   - \hat{\bm{e}}_z\partial_y A_x^*.
    \label{SynB}
\end{equation}
As a result, the position-dependent detuning $\delta(y,z,\theta)$ induces a synthetic magnetic field in the $y$-$z$ plane,
\begin{equation}
    \label{Bdirec}
    \bm{B} = B_0(0,-\cos\theta,\sin\theta),
\end{equation}
where $B_0 = \dfrac{2\eta k_0^2}{\Omega-\Omega_c}$. It is worth pointing out that the synthetic magnetic field $\bm{B}$ is always perpendicular to detuning gradient $\bm{\nabla}\delta$.

At zero temperature, the weakly interacting SOC BEC is well described by the GP equation
\begin{equation}
    i\frac{\partial}{\partial t} \psi = (H_0 +H_{\text{int}} ) \psi,
    \label{GPE}
\end{equation}
where $\psi=(\psi_1,\psi_2)^T$ represents the order parameter of the two components, satisfying the normalization condition $\int d\textbf{r}|\psi|^2 = N$, with $N$ denoting the total number of atoms. The mean-field interaction is given by 
$H_{\text{int}} = \text{diag}\left(g_{11}|\psi_1|^2 + g_{12}|\psi_2|^2,\  g_{12}|\psi_1|^2 + g_{22}|\psi_2|^2\right)$ where the coupling constants $g_{ij} = 4\pi\hbar^2 a_{ij}/m$ ($i,j \in \{1,2\}$) depend on the $s$-wave scattering lengths $a_{ij}$ and atomic mass $m$, \textcolor{black}{with $g_{11}, g_{22}$ and $g_{12} = g_{21}$ representing intra- and inter-species interactions, respectively}. For simplicity, we consider isotropic interactions, $g_{11} = g_{22} = g_{12} \equiv g$, which is a reasonable approximation for most alkali atoms.

% \section{Spinor Hydrodynamic theory}
To formulate HD equations, we parameterize the order parameters in the following form:
\begin{equation}
\label{eqwave}
    \binom{\psi_1}{\psi_2}=\binom{\sqrt{n_1}e^{i\phi_1}}{\sqrt{n_2}e^{i\phi_2}},
\end{equation}
where $n_j$ and $\phi_j$ denote the density and phase of $j$-th spin component. The Lagrangian density is expressed as
\begin{equation}
    \mathcal{L} = \sum_{j=1}^2 \frac{i\hbar}{2} \left( \psi_j^* \partial_t \psi_j - \psi_j \partial_t \psi_j^* \right) - \mathcal{E},
\end{equation}
where the energy density is defined as $\mathcal{E} = \psi^\dagger \left( H_0 + H_{\text{int}}/2 \right)\psi$. The properties of the system can be equally characterized by another new four variables: 
(i) the total density $n \equiv n_1 + n_2$;
(ii) the spin density $s_z \equiv n_1 - n_2$;
(iii) the total phase $\phi \equiv (\phi_1 + \phi_2)/2$; and 
(iv) the relative phase $\phi_R \equiv \phi_1 - \phi_2$. We are interested in the ground state and collective modes such as the dipole and quadrupole modes. The characteristic length scale of this model significantly exceeds the healing length; therefore, the quantum pressure term ($\propto \nabla^2\sqrt{n_j}$) can be neglected in our system. Applying the Euler-Lagrange equations to the four variational parameters, we derive the following set of closed equations of motion~\cite{PhysRevA.86.063621,PhysRevLett.118.145302,PhysRevLett.120.183202,PhysRevA.108.053316}:
\begin{subequations}
\label{eq:hydrodynamics}
\begin{align}
    \frac{\partial n}{\partial t} + \frac{\hbar}{m}\bm{\nabla}\cdot(n\bm{\nabla}\phi) - \frac{\hbar k_0}{m}\nabla_x s_z &= 0, \label{eqq1} \\
    \hbar\frac{\partial \phi}{\partial t} + \frac{\hbar^2}{2m}(\bm{\nabla}\phi)^2 + V_{\text{trap}} + gn - \frac{\Omega}{2}\frac{n}{\sqrt{n^2 - s_z^2}} &= 0, \label{eqq2} \\
    \frac{\partial s_z}{\partial t}+\frac{\hbar}{m}\bm{\nabla}\cdot(s_z\nabla\phi)-\frac{\hbar k_0}{m}\nabla_xn&=0, \label{eqq3}\\
    \frac{2\hbar^2 k_0}{m}\nabla_x\phi + \Omega\frac{s_z}{\sqrt{n^2 - s_z^2}} - 2\eta k_0(y\sin\theta + z\cos\theta) &= 0.\label{eqq4}
\end{align}
\end{subequations}
In our work, we focus on the zero-momentum phase, where the Raman coupling strength $\Omega$ exceeds the critical value $\Omega_c$. In this phase, where $\Omega$ is sufficiently large compared to low-energy excitations, the spin-orbit coupling explicitly breaks the $U(1)$ gauge symmetry associated with the relative phase $\phi_R$, dynamically locking it to zero ($\phi_R \to 0$)~\cite{PhysRevA.86.063621}. This symmetry breaking enables the elimination of $\phi_R$ from the HD description. In the absence of a detuning gradient, the ground state is spin balanced. A weak detuning gradient induces only small spin-density fluctuations, with $s_z \ll n$, allowing a perturbative treatment. After performing
a second-order Taylor expansion of the spin-dependent
terms in Eq.~\eqref{eqq4}, we obtain
\begin{equation}
\label{eqq6}
\frac{s_{z}}{n}=\frac{k_{0}}{\Omega}\left[\frac{2\hbar^{2}}{m}\nabla_{x}\phi+2\eta(y\sin\theta + z\cos\theta )\right].
\end{equation}
Substituting Eq.~\eqref{eqq6} into Eqs.~\eqref{eqq1} and \eqref{eqq2}, we eliminate the spin degree of freedom, leading to an effective spinor HD formulation governed by the total density $n$ and the total phase $\phi$:
\begin{widetext}
\begin{subequations}
\begin{align}
    \frac{\partial n}{\partial t}+\frac{\hbar}{m^{*}}\nabla_{x}(n\nabla_{x}\phi)+\frac{\hbar}{m}\nabla_{y}(n\nabla_{y}\phi)+\frac{\hbar}{m}\nabla_{z}(n\nabla_{z}\phi)-\frac{\Omega_{c}}{\hbar\Omega}\eta(\sin\theta y+\cos\theta z)\nabla_{x}n	&=0, \label{eq:reduced1} \\
    \hbar\frac{\partial\phi}{\partial t}+\frac{\hbar^{2}}{2m^{*}}(\nabla_{x}\phi)^{2}+\frac{\hbar^{2}}{2m}(\nabla_{y}\phi)^{2}+\frac{\hbar^{2}}{2m}(\nabla_{z}\phi)^{2}-\frac{\Omega}{2}+gn+V_{trap}-\frac{\Omega_{c}}{\Omega}\eta(\sin\theta y+\cos\theta z)\nabla_{x}\phi	&=0.\label{eq:reduced2}
\end{align}
\label{Hydrodynamic}
\end{subequations}
\end{widetext}
The HD equations~\eqref{Hydrodynamic} exhibit remarkable applicable ability—their time-dependent solutions capture dynamical evolution of SOC BEC, while the static limit ($\partial_t \to 0$) reveals equilibrium properties including the associated spin density configurations~\cite{PhysRevLett.120.183202,PhysRevA.110.043307}.

% Hydrodynamic theory offers a rigorous foundation for analyzing collective excitations in Bose-Einstein condensates through linear perturbation analysis. Within the Thomas-Fermi regime, it accurately describes various collective dynamics, including (but not limited to) dipole modes, quadrupole modes, and expansion dynamics, as demonstrated in \textcolor{red}{Refs}.~\cite{DALFOVO1997259,PhysRevLett.77.2360,PhysRevLett.83.4452,PhysRevLett.120.183202,PhysRevA.108.053316,PhysRevA.110.043307,PhysRevResearch.7.013219}.

\section{Ground States in a Synthetic magnetic Field with Tunable Orientation}\label{sec3}
% \subsection{Thomas Fermi Approximation}

%%%%%%%%%%%%%%%%%%%%%%%%%%%%%%
\begin{figure*}[t]
    \centerline{
    \includegraphics[width=1.0\textwidth]{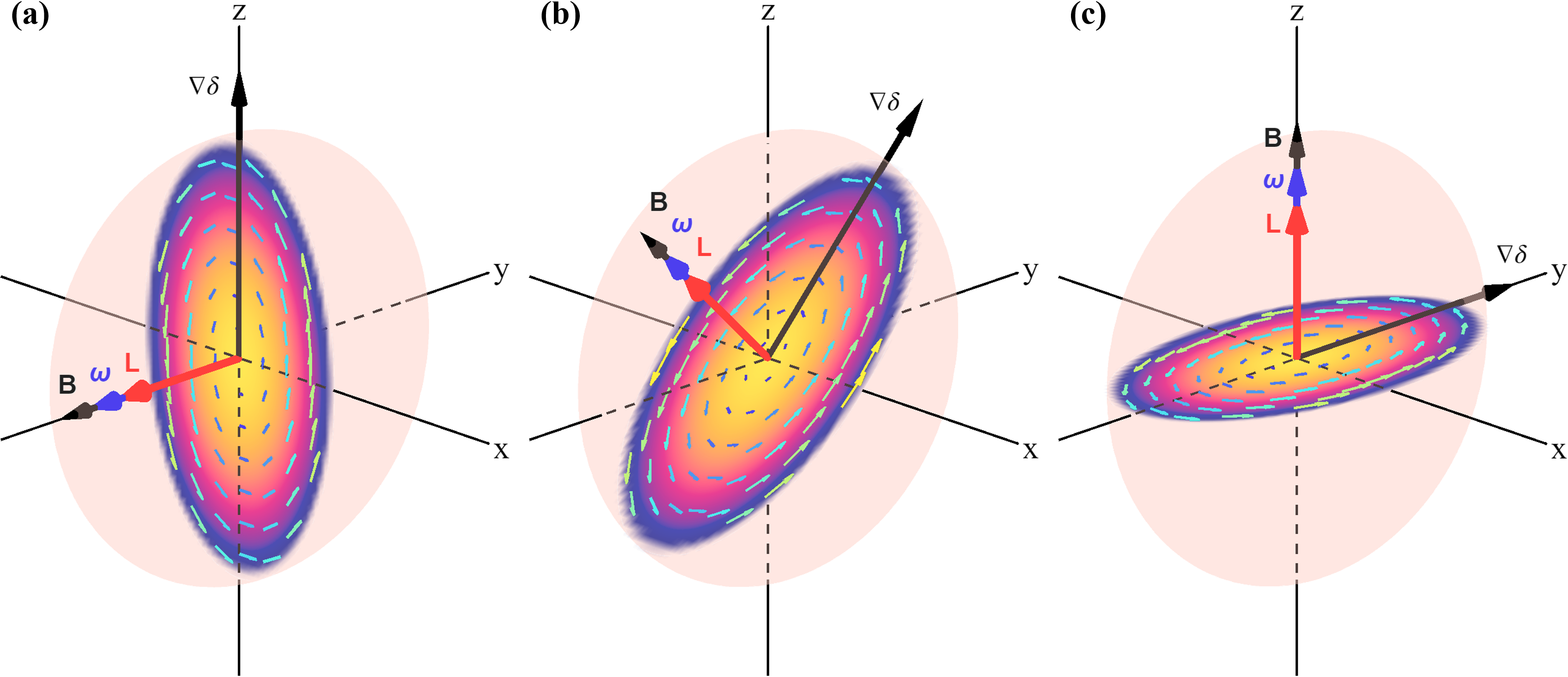}}
        \caption{
        \textcolor{black}{Velocity field distributions of the ground state for three different orientations of the synthetic magnetic field: (a) $\theta = 0^\circ$, (b) $45^\circ$, and (c) $90^\circ$. The position-dependent detuning $\delta(y,z,\theta)$ is characterized by its gradient $\bm{\nabla}\delta$ in a harmonically trapped BEC. This detuning induces the synthetic magnetic field $\bm{B}$, which is always perpendicular to both the detuning gradient $\bm{\nabla}\delta$ and the spin-orbit coupling direction, the latter being along the $x$-axis, as indicated by the two black arrows in each panel. The colored plane shown in each panel is perpendicular to the synthetic magnetic field $\bm{B}$, and is quantitatively described by the equation $-\beta_z y + \beta_y z = 0$. The color represents the local density of the condensate, while the arrows within the plane represent the magnitude and direction of the local velocity field. In addition to the synthetic magnetic field, the position-dependent detuning $\delta(y,z,\theta)$ also induces an angular velocity $\bm{\omega}$ (blue arrows) and an angular momentum $\bm{L}$ (red arrows). When the trapping frequencies satisfy $\omega_y = \omega_z$, both vectors are aligned with the synthetic magnetic field $\bm{B}$. However, when $\omega_y \ne \omega_z$, this alignment breaks down, as illustrated in Fig.~\ref{fig2}.} The outer pale pink contours delineate the effective boundary of BEC density profiles, a direct consequence of the confinement imposed by the harmonic trapping potential. The system parameters are configured with trapping frequencies $(\omega_x,\omega_y,\omega_z) = 2\pi\times(50\sqrt{3},50,50)$ Hz, detuning gradient $\eta = 0.001E_r$, Raman coupling strength $\Omega = 6E_r$ and particle number $N = 5\times10^4$.}
    \label{fig1}
\end{figure*}
%%%%%%%%%%%%%%%%%%%%%%%%%%%%%%

At equilibrium, the time-dependent part of order parameter can be expressed as, $\exp(-i\mu t/\hbar)$, where $\mu$ represents the chemical potential of the entire system. Substituting this expression into Eq.~\eqref{eqq2} and neglecting the second-order terms, Eq.~\eqref{eqq2} becomes
\begin{equation}
    -\mu + V_{\text{trap}} + gn - \frac{\Omega}{2} = 0.
\end{equation}
As a result, we obtain the Thomas-Fermi distribution of the ground state of the following form:
\begin{equation}
\label{ThomasFermi}
n_0 = \frac{\mu^\prime}{g}\left(1 - \frac{x^2}{R_x^2} - \frac{y^2}{R_y^2} - \frac{z^2}{R_z^2}\right),
\end{equation}
where $\mu' = \mu + \Omega/2$ and the Thomas-Fermi radii are defined as $R_\nu = \sqrt{2\mu'/\left(m\omega_\nu^2\right)}$ ($\nu=x,y,z$). The value of $\mu'$ is fixed by the normalization of the total density $n_0$ to the number of particle $N$, which yields~\cite{pethick2008bose}:
\begin{equation}
\mu' = \frac{1}{2}\hbar\bar{\omega}\left(\frac{15Na}{\bar{a}_{\text{ho}}}\right)^{2/5},
\end{equation}
where $\bar{a}_{\text{ho}} = \sqrt{\hbar/(m\bar{\omega})}$, $\bar{\omega} = \left(\omega_x\omega_y\omega_z\right)^{1/3}$ and $a=a_{ij}$ is the scattering length mentioned before.

\subsection{Velocity Field}

After carefully examining Eqs.~\eqref{eq:hydrodynamics}, we find that the phase of the ground state takes the polynomial form $\phi_0 = \alpha_{y} xy + \alpha_{z} xz$. Substituting $\phi_0$ into Eq.~\eqref{eqq6} yields a spin density distribution of the form $s_z = 2\left(\beta_y y + \beta_z z\right)n_0$. Inserting the equilibrium expressions $n_0$, $\phi_0$, and $s_z$ into Eqs.~\eqref{eq:hydrodynamics}, and neglecting the second-order terms related to the detuning gradient $\eta$, we obtain ground-state coefficients $\alpha_i$ and $\beta_i$ that explicitly depend on the orientation parameter $\theta$:
\begin{equation}  
\label{coefi}
\begin{aligned}  
\alpha_y &= 2\eta \frac{k_0^2}{\Omega} \frac{\omega_x^2}{\omega_{xy}^2} \sin\theta, \quad \beta_y = \eta \frac{k_0}{\Omega} \frac{\omega_x^2 + \omega_y^2}{\omega_{xy}^2} \sin\theta, \\  
\alpha_z &= 2\eta \frac{k_0^2}{\Omega} \frac{\omega_x^2}{\omega_{xz}^2} \cos\theta, \quad \beta_z = \eta \frac{k_0}{\Omega} \frac{\omega_x^2 + \omega_z^2}{\omega_{xz}^2} \cos\theta,
\end{aligned}  
\end{equation}
where the scissors mode frequencies of SOC BEC are defined as $\omega_{xy}=\sqrt{(m/m^{*})\omega_{x}^{2}+\omega_{y}^{2}}$ and $\omega_{xz}=\sqrt{(m/m^{*})\omega_{x}^{2}+\omega_{z}^{2}}$ without the detuning gradient.

Comparing Eq.~\eqref{eqq1} to the continuity equation, $ \partial_t n + \bm{\nabla}\cdot\bm{j}=0$, we find that the current density $\bm{j}$~\cite{PhysRevA.86.063621} can be naturally separated into a canonical term and a spin-related term. Using the standard relation between the velocity field and current density, $\bm{v} = \bm{j} / n$, and substituting the phase $\phi = \phi_0$ and the spin density $s_z$ into $\bm{j}$, we obtain the explicit form of the velocity field:
\begin{equation}
\label{velo}
\begin{aligned}
    \bm{v} &= \bm{v}^c + \bm{v}^s, \\ 
    \bm{v}^c &= \frac{\hbar}{m} \left( \alpha_y y + \alpha_z z, \, \alpha_y x, \, \alpha_z x \right), \\ 
    \bm{v}^s &= -\frac{2\hbar k_0}{m} \left( \beta_y y + \beta_z z, \, 0, \, 0 \right),
\end{aligned}
\end{equation}
where $\bm{v}^c$ and $\bm{v}^s$ correspond to the canonical and spin-related velocity fields, respectively.

Spin-orbit coupling profoundly affects the rotational properties of BECs, breaking irrotationality. The position-dependent detuning $\delta(y,z,\theta)$ drives rotational behavior in the condensate's ground state~\cite{PhysRevLett.118.145302,PhysRevLett.120.183202}. Figure~\ref{fig1} shows the ground-state velocity field distributions for three orientations ($\theta = 0^\circ,45^\circ,90^\circ$) of the synthetic magnetic field. Black arrows indicate the directions of the synthetic magnetic field $\bm{B}$ and the detuning gradient $\bm{\nabla}\delta$, which are mutually perpendicular. The orientation parameter $\theta$ corresponds to the angle between $\bm{B}$ and $-y$-axis. The rotational velocity field and synthetic magnetic field induce angular velocity $\bm{\omega}$ (blue arrows) and angular momentum $\bm{L}$ (red arrows), as shown in Fig.~\ref{fig1}. In all illustrated cases where the trapping frequencies all satisfy $\omega_y = \omega_z$, the three vectors $\bm{B}$, $\bm{\omega}$ and $\bm{L}$ are perpendicular to the elliptical velocity field plane (colored regions). This parallelism persists regardless of the orientation angle $\theta$ of the synthetic magnetic field. In contrast, if $\omega_y \neq \omega_z$, the three vectors remain mutually parallel only at specific orientations, namely $\theta = 0^\circ$ or $\theta = 90^\circ$. We now examine in detail the relationships among these induced vectors.

%%%%%%%%%%%%%%%%%%%%%%%%%%%%%%
\begin{figure*}[t]
    \centerline{
    \includegraphics[width=1\textwidth]{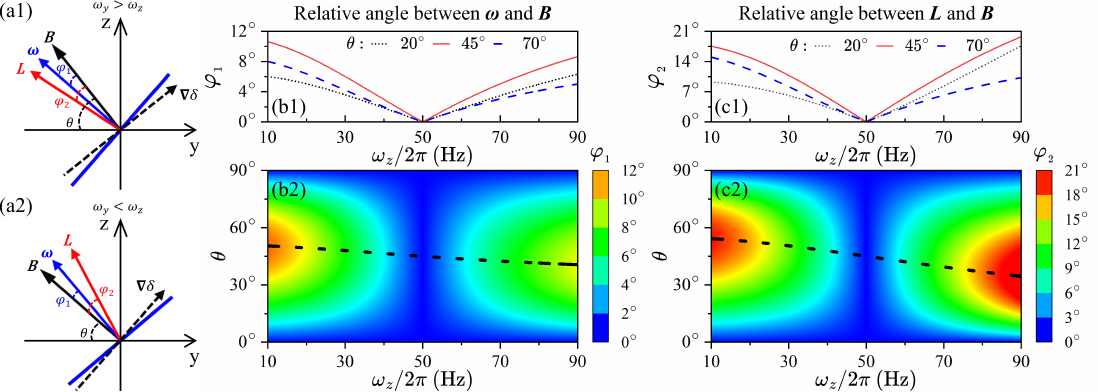}}
        \caption{\textcolor{black}{Schematic of two distinct types of angular misalignment, $\varphi_1$ and $\varphi_2$, arising from the combined effects of position-dependent detuning and trap anisotropy. (a)~$\varphi_1$ is the angle between the angular velocity $\bm{\omega}$ (blue arrows) and the synthetic magnetic field $\bm{B}$ (black solid arrows), while $\varphi_2$ is the angle between the angular momentum $\bm{L}$ (red arrows) and $\bm{B}$. Under general conditions where the orientation parameter $\theta \ne 0^\circ$ or $90^\circ$, and trap anisotropy is present (i.e., $\omega_y \ne \omega_z$), the vectors $\bm{B}$, $\bm{\omega}$, and $\bm{L}$ are no longer mutually parallel. In both cases, $\bm{\omega}$ remains perpendicular to the representative velocity field plane (blue solid lines), and $\bm{B}$ remains perpendicular to the detuning gradient $\bm{\nabla}\delta$ (black dashed arrows). (a1) and (a2) represent the $\omega_y>\omega_z$ and $\omega_y>\omega_z$, respectively. (b1)~Misalignment angle $\varphi_1$ as a function of $\omega_z$, with fixed $(\omega_x, \omega_y)$ and Raman coupling $\Omega$, shown for three representative orientation parameters: $\theta = 20^\circ$ (black dotted), $45^\circ$ (red solid), and $70^\circ$ (blue dashed). (b2)~Colormap of $\varphi_1$ as a function of $\omega_z$ and $\theta$. Color indicates the magnitude of $\varphi_1$; the black dashed curve marks the locus of maximum misalignment, determined by $\theta = \arccos[(1 - \gamma_1)^{-1}]$. (c1)~Misalignment angle $\varphi_2$ between angular momentum $\bm{L}$ and synthetic magnetic field $\bm{B}$, shown under the same parameters as in (b1). (c2)~Colormap of $\varphi_2$ as a function of $\omega_z$ and $\theta$. The qualitative behavior resembles that of $\varphi_1$, with noticeable quantitative differences in the curve ranges and color distribution compared to (b1) and (b2). The black dashed curve marks the condition $\theta = \arccos[(1 - \gamma_2)^{-1}]$ for maximum misalignment. The other system parameters are set as follows: trapping frequencies $(\omega_x, \omega_y) = 2\pi \times (50\sqrt{3}, 50) $ Hz, detuning gradient $\eta = 0.001E_r$, and Raman coupling strength $\Omega = 6E_r$.}}
    \label{fig2}
\end{figure*}
%%%%%%%%%%%%%%%%%%%%%%%%%%%%%%

\subsection{Angular Velocity}

As shown in Eq.~\eqref{velo}, the velocity field of SOC BEC with a detuning gradient exhibits a structure analogous to the rotational velocity field of a classical rigid body, reflecting the system's underlying rotational characteristics. In addition to the rotational behavior, the specific form of the velocity field is determined by the trap geometry and exhibits some deformation compared to perfect rigid rotation. To quantify the local rotational and deformation features of the ground state, one can introduce the velocity gradient tensor~\cite{Gurtin2010TheMA}: 
\begin{equation}
    \begin{aligned}
        \bm{\nabla}\bm{v}&=\begin{pmatrix}\nabla_x v_{x} & \nabla_y v_{x} & \nabla_z v_{x}\\
        \nabla_x v_{y} & \nabla_y v_{y} & \nabla_z v_{y}\\
        \nabla_x v_{z} & \nabla_y v_{z} & \nabla_z v_{z}
        \end{pmatrix}.
    \end{aligned}
\end{equation}

The symmetric part of $\bm{\nabla}\bm{v}$, the so-called strain-rate tensor $\mathbf{D}$, describes shear and expansion: 
\begin{equation}
    \begin{aligned}
        \mathbf{D} &= \frac{1}{2} \left[\bm{\nabla} \bm{v} + (\bm{\nabla} \bm{v})^{T}\right]\\
        &=\frac{\hbar}{m}\begin{pmatrix}0 & \alpha_{y}-k_{0}\beta_{y} & \alpha_{z}-k_{0}\beta_{z}\\
        \alpha_{y}-k_{0}\beta_{y} & 0 & 0\\
        \alpha_{z}-k_{0}\beta_{z} & 0 & 0
        \end{pmatrix}.
    \end{aligned}
\end{equation}
In the ground state, the trace of $\mathbf{D}$, which quantifies the volumetric expansion rate, vanishes: $\text{Tr}(\mathbf{D}) = \bm{\nabla}\cdot\bm{v} = 0$. Moreover, all three diagonal components $D_{ii}\ (i = x, y, z)$ are individually zero. This indicates that the condensate exhibits no radial flow, maintaining local incompressibility and volume conservation, resembling the behavior of an incompressible classical fluid. In contrast, the off-diagonal components of $\mathbf{D}$, such as $D_{xy} = D_{yx} = \dfrac{\hbar}{m}(\alpha_y - k_0 \beta_y)$ and $D_{xz} = D_{zx} = \dfrac{\hbar}{m}(\alpha_z - k_0 \beta_z)$, represent shear deformation rates in the $x$-$y$ and $x$-$z$ planes, respectively. These nonzero components are responsible for generating elliptical streamlines in the velocity field when the trapping frequencies are anisotropic.

The antisymmetric part of the velocity gradient tensor describes rotational behavior:
\begin{equation}
    \begin{aligned}
        \mathbf{W} &= \frac{1}{2} \left[\bm{\nabla} \bm{v} - (\bm{\nabla} \bm{v})^{T}\right]\\
        &=\frac{\hbar k_{0}}{m}\begin{pmatrix}0 & -\beta_{y} & -\beta_{z}\\
        \beta_{y} & 0 & 0\\
        \beta_{z} & 0 & 0
        \end{pmatrix}.
    \end{aligned}
\end{equation}
The local angular velocity is related to $\mathbf{W}$, through the relation $W_{ij} = -\epsilon_{ijk} \omega_k$, where $\epsilon_{ijk}$ denotes the Levi-Civita symbol, and it can be calculated by
\begin{equation}
    \label{omegaveo}
    \begin{aligned}
        \bm{\omega} &= \frac{1}{2}\bm{\nabla}\times \bm{v} \\
        &=\frac{\hbar k_0}{m} (0, -\beta_z, \beta_y),
    \end{aligned}
\end{equation}
which quantitatively characterizes the rigid-like rotation velocity field shown in Eq.~\eqref{velo}. 

A particularly symmetric configuration arises in the case of a spherical harmonic trap, i.e., $\omega_x = \omega_y = \omega_z$, where the coefficients satisfy $\alpha_y = k_0 \beta_y$ and $\alpha_z = k_0 \beta_z$ simultaneously. In this case, the strain-rate tensor vanishes entirely ($\mathbf{D} = 0$), and the velocity field becomes $\bm{v} = \bm{\omega} \times \bm{r}$, representing a pure rigid-body rotation. More generally, when the trap is anisotropic, both shear (via $\mathbf{D}$) and rotation (via $\mathbf{W}$) coexist and the resulting flow exhibits elliptical, closed streamlines, which arise from the balance between deformation and rotation. When the three dipole frequencies are degenerate, i.e., $\sqrt{m/m^*}\omega_x = \omega_y=\omega_z$, the angular velocity $\bm{\omega}$ is parallel to the synthetic magnetic field $\bm{B}$, with an elliptical flow, as shown in Fig.~\ref{fig1}. In the most general case, where $\omega_x \ne \omega_y \ne \omega_z$, $\bm{\omega}$ is not necessarily parallel to $\bm{B}$.

To quantify the misalignment between the angular velocity $\bm{\omega}$ and the synthetic magnetic field $\bm{B}$ in the general case, we compare Eqs.~\eqref{Bdirec} and \eqref{omegaveo}. This yields an explicit expression for the angle $\varphi_1$ between the two vectors:
\begin{equation}
\label{misalignment1}
\cos\varphi_1 = \frac{\gamma_1 \cos^2\theta + \sin^2\theta}{\sqrt{\gamma_1^2 \cos^2\theta + \sin^2\theta}},
\end{equation}
where the coefficient $\gamma_1 = \dfrac{(\omega_x^2 + \omega_z^2)\omega_{xy}^2}{(\omega_x^2 + \omega_y^2)\omega_{xz}^2}$ includes all three trap frequencies $(\omega_x, \omega_y, \omega_z)$, and depends on the scissors mode frequencies $\omega_{xy}$ and $\omega_{xz}$, both of which are modified by the spin-orbit coupling. Consequently, Eq.~\eqref{misalignment1} reveals how the trapping potential and spin-orbit coupling affect the direction of the angular velocity relative to the synthetic magnetic field. To visualize the misalignment, we present in Fig.~\ref{fig2}\hyperref[fig2]{(b2)} a color map of the misalignment angle $\varphi_1$ as a function of the trapping frequency $\omega_z$ and the orientation $\theta$ of the synthetic magnetic field. In this map, the color intensity indicates the magnitude of the angular deviation $\varphi_1$.

Perfect alignment ($\varphi_1 = 0^\circ$) occurs when $\gamma_1 = 1$, which can be achieved under two distinct conditions. The first is when the trap is axially symmetric, i.e., the frequencies satisfy $\omega_y = \omega_z$. This corresponds to $\omega_z = \omega_y = 2\pi\times50$ Hz in Fig.~\ref{fig2}\hyperref[fig2]{(b2)}, and is exemplified by the aligned black and blue arrows in Fig.~\ref{fig1}\hyperref[fig1]{(b)}. The second is when the synthetic magnetic field is aligned with one of the symmetry axes of the harmonic trap, namely $\theta = 0^\circ$ or $90^\circ$, in which case alignment occurs regardless of trap anisotropy. Two such configurations are illustrated in Fig.~\ref{fig1}\hyperref[fig1]{(a)} and \hyperref[fig1]{(c)}.

In Figs.~\ref{fig2}\hyperref[fig2]{(a1)} and \hyperref[fig2]{(a2)}, we illustrate two representative scenarios under general conditions with $\theta \ne 0^\circ,90^\circ$. Whether the two vectors $\bm{\omega}$ and $\bm{B}$ align with each other depends on the trap geometry. When $\omega_y > \omega_z$ (i.e., $\gamma_1 > 1$), the angular velocity $\bm{\omega}$ tilts toward the $-y$-axis [Fig.~\ref{fig2}\hyperref[fig2]{(a1)}]; conversely, when $\omega_y < \omega_z$ (i.e., $\gamma_1 < 1$), it tilts toward the $+z$-axis [Fig.~\ref{fig2}\hyperref[fig2]{(a2)}]. The black dashed curve in Fig.~\ref{fig2}\hyperref[fig2]{(b2)} traces the position of the maximum misalignment angle $\varphi_1^\text{max}$ for a given $\omega_z$, and is quantitatively described by
\begin{equation}
    \theta = \arccos{\left[(1 - \gamma_1)^{-1}\right]}.
\end{equation}

% Notably, under the chosen parameters in Fig.~\ref{fig2}, the appearance of misalignment is not symmetric about the axis $\omega_z = \omega_y = 2\pi \times 50$ Hz, as evidenced by the three variations in Fig.~\ref{fig2}\hyperref[fig2]{(b1)} for different orientations of the synthetic magnetic field. This asymmetry is further visualized in the color map of Fig.~\ref{fig2}\hyperref[fig2]{(b2)}, where the misalignment angle $\varphi_1$ is encoded by the color intensity.

These observations demonstrate that the direction of the angular velocity is geometrically determined by the anisotropy of the trapping potential and the direction of the detuning gradient $\bm{\nabla}\delta$.

\subsection{Angular Momentum}

When the system develops a rigid rotational velocity field, it naturally exhibits an associated angular momentum. The angular momentum operator are defined as:
\begin{equation}
    \hat{\bm{L}} = \hat{\bm{r}}\times \hat{\bm{p}}.
\end{equation}
In the presence of spin-orbit coupling, the momentum acquires a spin-dependent contribution along the $x$-direction, given by $-\hbar k_0 \sigma_z\hat{\bm{e}}_x$. As a result, the angular momentum can be naturally separated into canonical components ($\hat{L}_{x}^c, \hat{L}_{y}^c, \hat{L}_{z}^c$) and spin components ($\hat{L}_{y}^s, \hat{L}_{z}^s$). Using the Thomas-Fermi density profile $n_0$ and the ground-state phase ansatz $\phi_0$, we compute the expectation values of the angular momentum components 
\begin{equation}
    \label{angularcomp}
    \begin{aligned}
        \langle\hat{L}_{x}^{c}\rangle &= \langle y\hat{p}_z - z\hat{p}_y\rangle = 0,\\
        \langle\hat{L}_{y}^{c}\rangle &= \langle z\hat{p}_x - x\hat{p}_z\rangle = \hbar \frac{R_{z}^{2}-R_{x}^{2}}{7}\alpha_{z},\\
        \langle\hat{L}_{z}^{c}\rangle &= \langle x\hat{p}_y - y\hat{p}_x\rangle = \hbar\frac{R_{x}^{2}-R_{y}^{2}}{7}\alpha_{y},\\
        \langle\hat{L}_{y}^{s}\rangle &= \langle -\hbar k_0 z\sigma_z\rangle = -\hbar\frac{2R_{z}^{2}}{7}k_{0}\beta_{z},\\
        \langle\hat{L}_{z}^{s}\rangle &= \langle \hbar k_0 y\sigma_z\rangle = \hbar\frac{2R_{y}^{2}}{7}k_{0}\beta_{y},
    \end{aligned}
\end{equation}
where $\langle \hat{\mathcal{O}}\rangle=\int[\mathcal{O}n(x,y,z)]d\mathbf{r}$ denotes the expectation value of the operator $\hat{\mathcal{O}}$. From Eq.~\eqref{angularcomp}, it is evident that the angular momentum $\bm{L}$ has no $x$-component and thus lies entirely within the $y$-$z$ plane. Using the expression for the Thomas-Fermi radii $R_\nu$ in Eq.~\eqref{ThomasFermi}, the resulting vector $\bm{L}$ can be expressed as
\begin{equation}
    \label{angularmo}
    \begin{aligned}
            \bm{L} &= \left(\langle\hat{L}_{x}^c\rangle,\langle\hat{L}_{y}^c\rangle+\langle\hat{L}_{y}^s\rangle,\langle\hat{L}_{z}^c\rangle+\langle\hat{L}_{z}^s\rangle\right)\\
            &=\eta\frac{4\Omega_c\mu^{\prime}}{7\Omega\hbar}\left(0,-\frac{\cos\theta}{\omega_{xz}^{2}},\frac{\sin\theta}{\omega_{xy}^{2}}\right),
    \end{aligned}
\end{equation}
where the direction of $\bm{L}$ is determined by the frequencies of the scissors modes in the $x$-$z$ and $y$-$z$ planes, as well as the orientation parameter $\theta$. Analogous to the analysis of the misalignment angle $\varphi_1$ between the angular velocity $\bm{\omega}$ and the synthetic magnetic field $\bm{B}$, the misalignment angle $\varphi_2$ between $\bm{L}$ and $\bm{B}$ can be derived by comparing Eqs.~\eqref{Bdirec} and \eqref{angularmo}. The resulting expression is
\begin{equation}
    \cos\varphi_{2} = \frac{\gamma_{2}\cos^{2}\theta + \sin^{2}\theta}{\sqrt{\gamma_{2}^{2}\cos^{2}\theta + \sin^{2}\theta}},
\end{equation}
where the coefficient $\gamma_2 = \omega_{xy}^2 / \omega_{xz}^2$ is determined by the frequencies of two scissors modes and it is always not equal to $\gamma_1$.

As shown in Figs.~\ref{fig2}\hyperref[fig2]{(a)} and \hyperref[fig2]{(c)}, the qualitative behavior of the misalignment angle $\varphi_2$ closely follows that of $\varphi_1$, reflecting the underlying geometric relation between the angular momentum and the synthetic magnetic field. For example, in Figs.~\ref{fig2}\hyperref[fig2]{(a1)} and \hyperref[fig2]{(a2)}, the angular momentum (red arrows) exhibit a larger misalignment angle $\varphi_2$ compared to $\varphi_1$. The quantitative differences between $\varphi_1$ and $\varphi_2$ are further illustrated in Figs.~\ref{fig2}\hyperref[fig2]{(b)} and \hyperref[fig2]{(c)}. The black dashed curve in Fig.~\ref{fig2}\hyperref[fig2]{(c2)} marks the position of the maximum misalignment angle $\varphi_2^\text{max}$, given by $\theta = \arccos[(1 - \gamma_2)^{-1}]$.

\begin{figure}[t]
    \centerline{
    \includegraphics[width=0.45\textwidth]{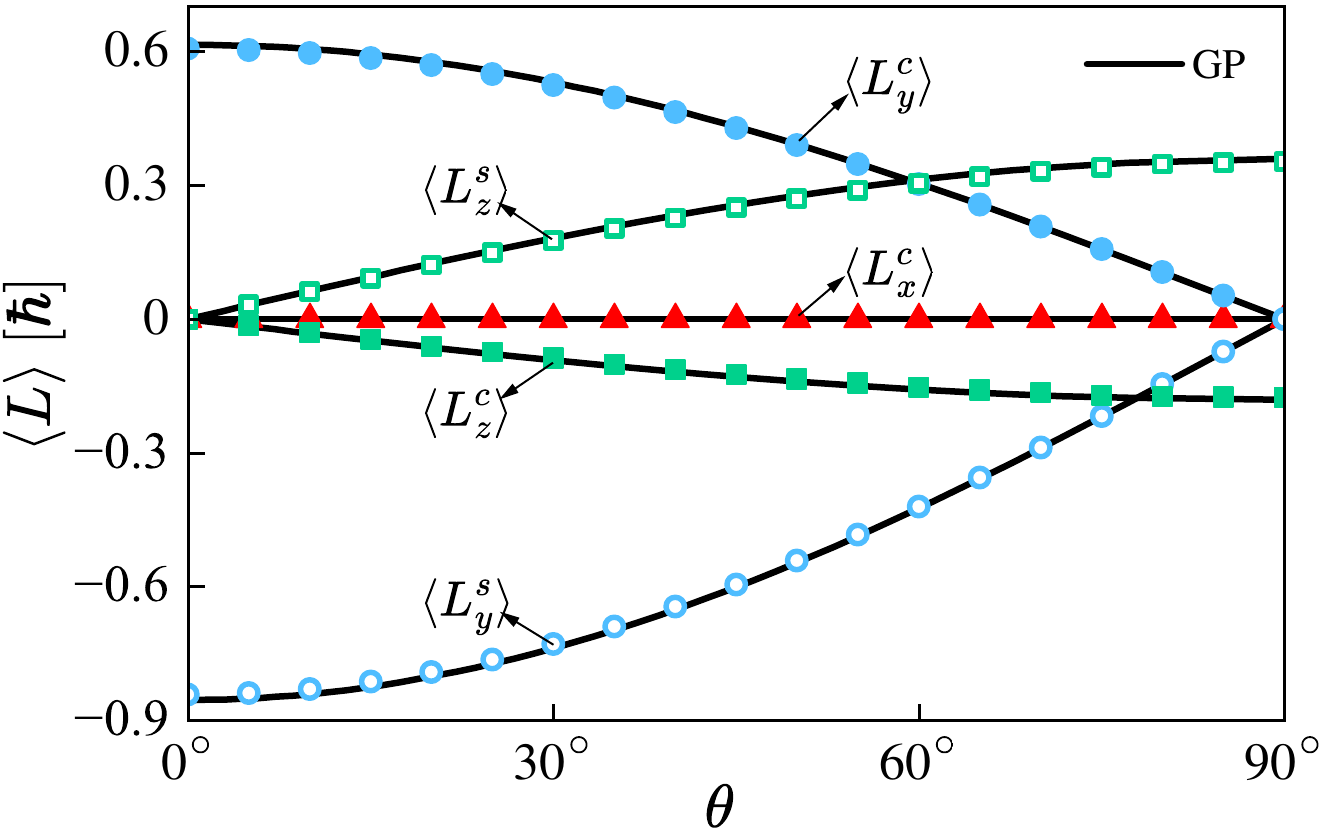}}
        \caption{Angular momentum components of the ground state for different orientations of the synthetic magnetic field in an anisotropic trap. Solid black curves show results from GP simulations. Colored markers indicate predictions from HD theory: solid blue spheres, hollow blue spheres, solid green squares, hollow green squares, and red triangles represent five distinct angular momentum components. System parameters are: particle number $N = 5\times10^4$, trapping frequencies $(\omega_x,\omega_y,\omega_z) = 2\pi \times (50\sqrt{3}, 50, 35)$ Hz, detuning gradient $\eta = 0.001E_r$, and Raman coupling strength $\Omega = 6E_r$.}
    \label{fig3}
\end{figure}

Figure~\ref{fig3} illustrates the dependence of the angular momentum components on the trap geometry for an anisotropic harmonic trap with frequencies $(\omega_x, \omega_y, \omega_z) = 2\pi \times (50\sqrt{3}, 50, 35)$ Hz. The results from GP simulations are found to be in excellent agreement with the predictions of HD theory.

%%%%%%%%%%%%%%%%%%%%%%%%%%%%%%
%%%%%%%%%%%%%%%%%%%%%%%%%%%%%%

\section{Quench Dynamics}\label{sec4}

The previous section establishes a spinor HD description of the system's ground state, characterizing the spatial relations among the synthetic magnetic field, velocity field, angular velocity, and angular momentum in different system geometry. In this section, we investigate the collective dynamics in the regime where the dipole frequencies are degenerate in the absence of a detuning gradient, i.e., $\sqrt{m/m^*}\omega_x = \omega_y = \omega_z$. Under this condition, the coupling between these dipole modes exhibits interesting beat effects. Furthermore, the scissors modes also become degenerate, with frequencies satisfying $\omega_{xy} = \omega_{xz} = \omega_{yz}$, and exhibit similar beating effects. We adopt the parameter set presented in Fig.~\ref{fig1}, with trapping frequencies $(\omega_x, \omega_y, \omega_z) = 2\pi \times (50\sqrt{3}, 50, 50)$ Hz, Raman coupling $\Omega = 6 E_r$, and detuning gradient $\eta = 0.001 E_r$. 

%%%%%%%%%%%%%%%%%%%%%%%%%%%%%%
\begin{figure}[b]
    \centerline{
    \includegraphics[width=0.42\textwidth]{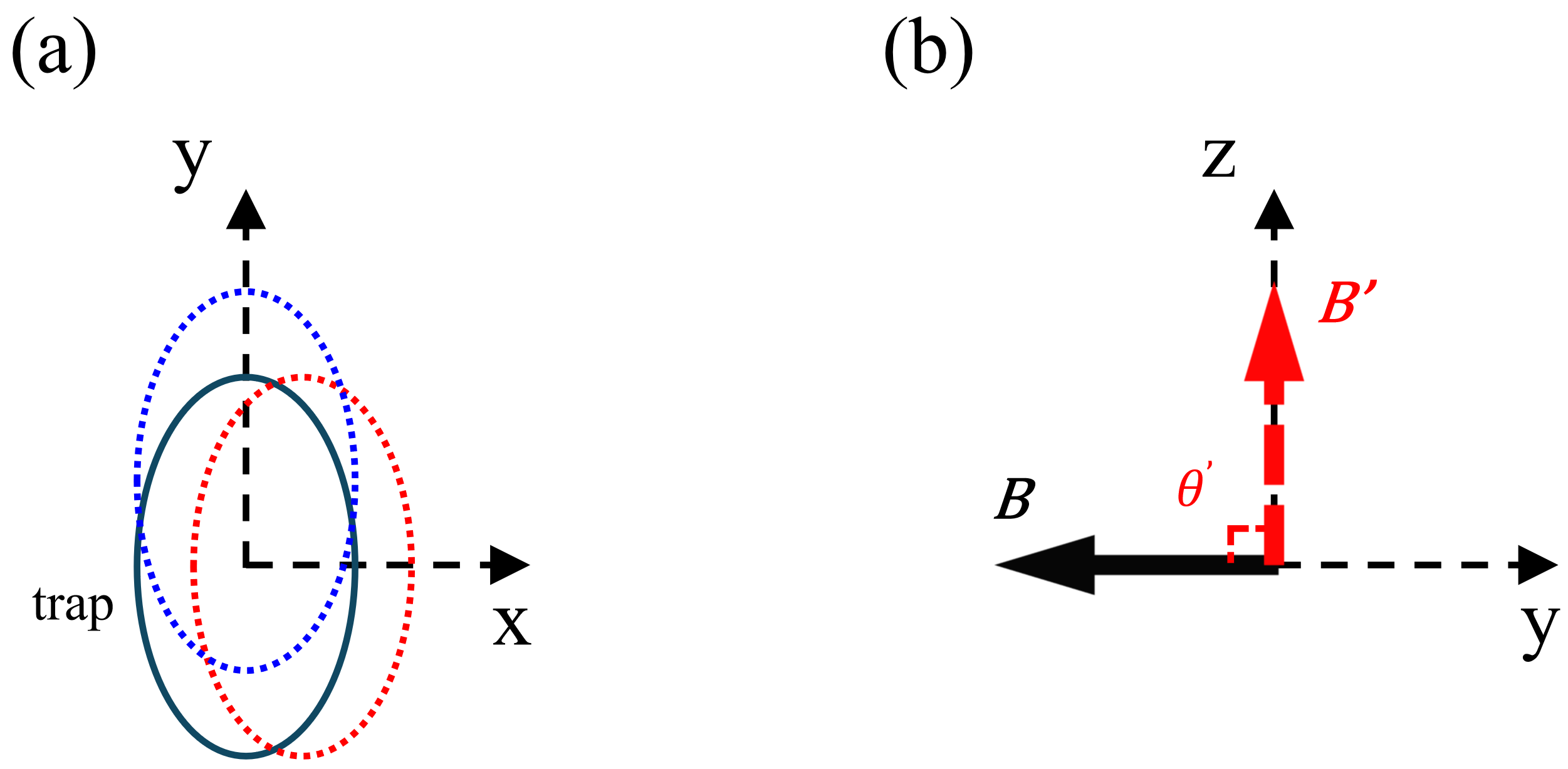}}
        \caption{Depiction of quench protocols used to selectively excite dipole and quadrupole modes. (a)~Top view of the trapping potential in the $x$-$y$ plane for a harmonic trap with $(\omega_x, \omega_y, \omega_z) = 2\pi \times (50\sqrt{3}, 50, 50)$ Hz and Raman coupling strength $\Omega = 6E_r$. The black solid circle marks the equilibrium trap center; red and blue dashed circles indicate displaced positions along the $x$ and $y$ directions, respectively, used to excite dipole oscillations. (b)~Side view in the $y$-$z$ plane showing the orientation of the synthetic magnetic field. The black arrow indicates the initial direction ($\theta_0 = 0^\circ$, along $-y$), while the red dashed arrow shows the quenched orientation ($\theta' = 90^\circ$, along $+z$), used to induce quadrupole-mode dynamics.}
    \label{fig4}
    \end{figure}
%%%%%%%%%%%%%%%%%%%%%%%%%%%%%%

To systematically explore how the orientation of the synthetic magnetic field affects dipole and quadrupole modes, we employ two distinct quenching protocols through HD and GP simulations:

1. An abrupt displacement of the trap center, 
\begin{equation}
    V_\text{trap}(\mathbf{r}) \xrightarrow{\text{quench}} V_\text{trap}(\mathbf{r}-\mathbf{r}_0),
\end{equation}
which selectively excites dipole-mode dynamics, as illustrated in Fig.~\ref{fig4}\hyperref[fig4]{(a)}.

2. A sudden rotation of the orientation of the synthetic magnetic field $\bm{B}$, 
\begin{equation}
    \theta_0 \xrightarrow{\text{quench}} \theta',
\end{equation} 
simultaneously exciting scissors modes and quadrupole modes, as illustrated in Fig.~\ref{fig4}\hyperref[fig4]{(b)}.

To analyze the collective excitation of the system, we first consider small perturbations around the equilibrium state. By introducing the ansatz $n = n_0 + \delta n$ and $\phi = \phi_0 + \delta\phi$ into the HD equations~\eqref{Hydrodynamic}, we can derive the linearized equations governing the dynamics of the density variation $\delta n$ and phase variation $\delta\phi$. In this process, we neglect the quadratic terms of the coefficients $\alpha_y$ and $\alpha_z$, as well as their cross terms, which arise from the phase gradient contribution in Eq.~\eqref{eq:reduced2}. This is justified as both $\alpha_y$ and $\alpha_z$ are proportional to the small detuning gradient $\eta = 0.001E_r$, making their second-order effects negligible. The linearized equations for the perturbation dynamics are:
\begin{widetext}
\begin{equation}
\label{perturb}
\begin{aligned}
\frac{\partial \delta n}{\partial t} &+ \frac{\hbar}{m^*}\nabla_x(n_0 \nabla_x \delta\phi) + \frac{\hbar}{m}\left[\nabla_y(n_0 \nabla_y \delta\phi) + \nabla_z(n_0 \nabla_z \delta\phi)\right]
- (\omega_{\text{eff,1}}y + \omega_{\text{eff,2}}z)\nabla_x \delta n + \omega_{\text{eff,1}}'x\nabla_y \delta n + \omega_{\text{eff,2}}'x\nabla_z \delta n = 0, \\
\frac{\partial \delta\phi}{\partial t} &+ \frac{g}{\hbar}\delta n - (\omega_{\text{eff,1}}y + \omega_{\text{eff,2}}z)\nabla_x \delta\phi + \omega_{\text{eff,1}}'x\nabla_y \delta\phi + \omega_{\text{eff,2}}'x\nabla_z \delta\phi = 0,
\end{aligned}
\end{equation}
\end{widetext}
where the effective frequencies $\omega_{\text{eff,i}}$ and $\omega_{\text{eff,i}}^\prime$ \textcolor{black}{($i = 1,2$)} \textcolor{black}{depend on} the spin-orbit coupling and detuning parameters:
\begin{equation}
\begin{aligned}
\omega_{\text{eff,1}} &= \frac{\eta\Omega_c}{\hbar\Omega}\frac{\omega_y^2}{\omega_{xy}^2}\sin\theta, \quad \omega_{\text{eff,2}} = \frac{\eta\Omega_c}{\hbar\Omega}\frac{\omega_z^2}{\omega_{xz}^2}\cos\theta, \\
\omega_{\text{eff,1}}' &= \frac{\eta\Omega_c}{\hbar\Omega}\frac{\omega_x^2}{\omega_{xy}^2}\sin\theta, \quad \omega_{\text{eff,2}}' = \frac{\eta\Omega_c}{\hbar\Omega}\frac{\omega_x^2}{\omega_{xz}^2}\cos\theta.
\end{aligned}
\end{equation}

\subsection{Dipole Oscillation Following a Quench of the Trap Center}

%%%%%%%%%%%%%%%%%%%%%%%%%%%%%%
\begin{figure*}[t]
    \centerline{
    \includegraphics[width=1\textwidth]{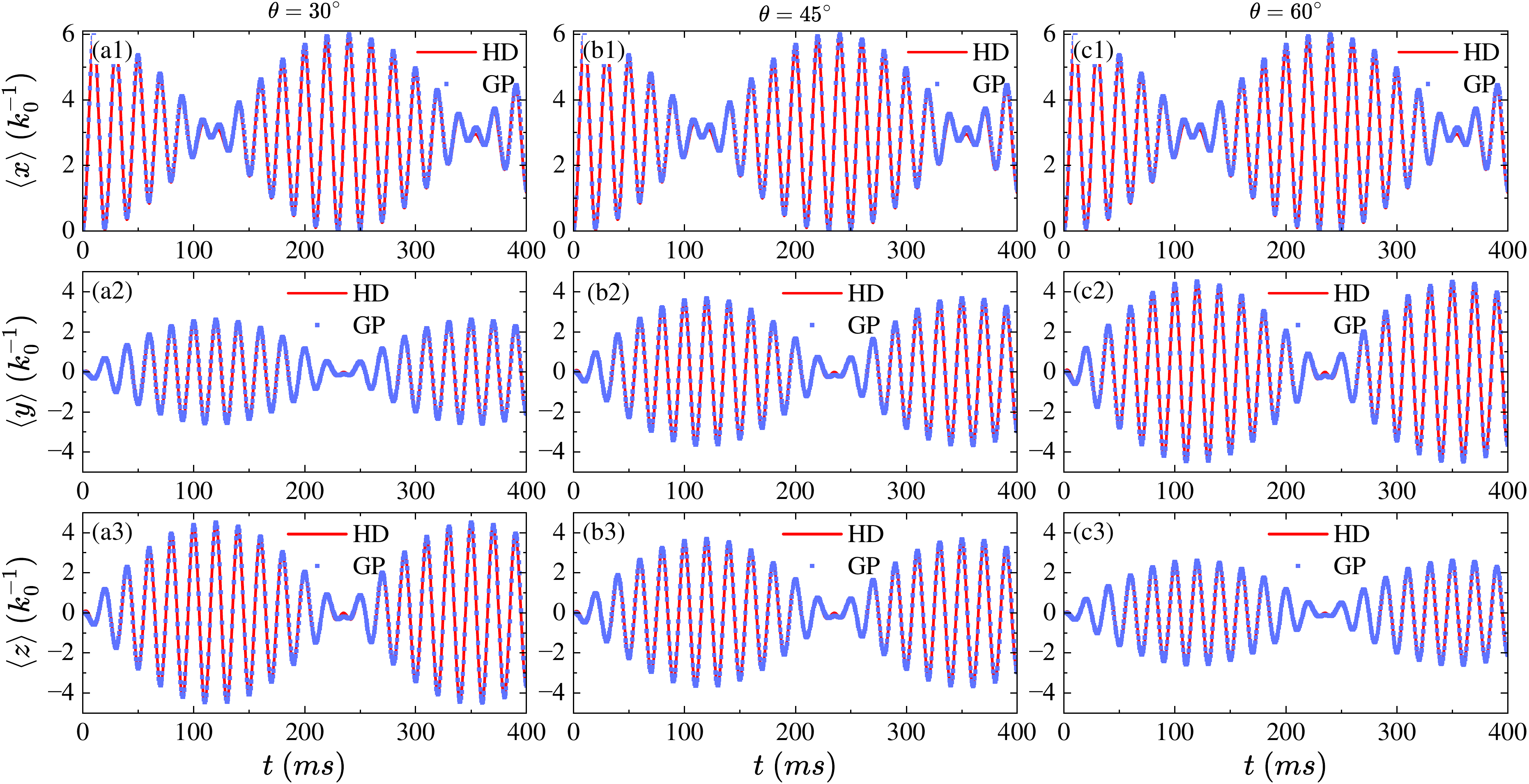}}
        \caption{Time evolution of the dipole-mode excitation along the $x$-direction, illustrating a clear beating effect for various orientations of the synthetic magnetic field. (a)-(c)~correspond to $\theta = 30^\circ$, $45^\circ$, and $60^\circ$, respectively. Red curves represent HD theory predictions, while blue dots denote results from GP simulations. The system parameters are: trapping frequencies $(\omega_x, \omega_y, \omega_z) = 2\pi \times (50\sqrt{3}, 50, 50)$ Hz, detuning gradient $\eta = 0.001 E_r$, Raman coupling strength $\Omega = 6 E_r$, and particle number $N = 5 \times 10^4$. Initial perturbations differ between methods: GP simulations implement a spatial displacement $x_0 = 3 k_0^{-1}$ along the $x$-axis, whereas the HD theory uses the initial condition $\epsilon_1 = 3 \left(\frac{4}{15} \pi R_x^2 R_y R_z \right)^{-1}$ as given in Eq.~\eqref{ansatzdipole}.}
    \label{fig5}
\end{figure*}
%%%%%%%%%%%%%%%%%%%%%%%%%%%%%%

%%%%%%%%%%%%%%%%%%%%%%%%%%%%%%
\begin{figure*}[t]
    \centerline{
    \includegraphics[width=1\textwidth]{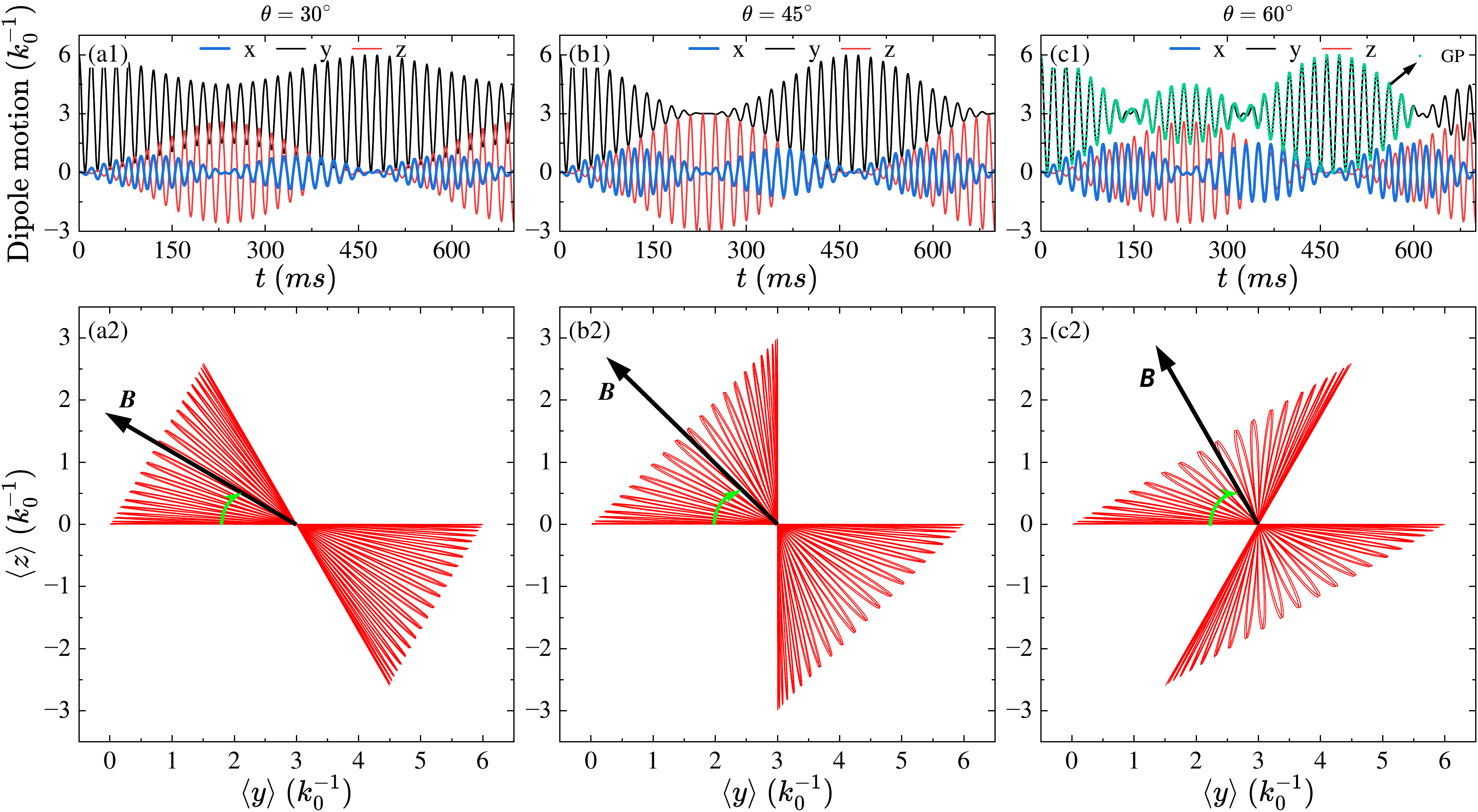}}
        \caption{Quenched $y$-dipole mode dynamics under the various orientations of the synthetic magnetic field. (a1)-(c1)~Temporal beating patterns of dipole oscillations along the $x$ (blue), $y$ (black), and $z$ (red) axes. Green dots in panel (c1) indicate the results from GP simulations. (a2)-(c2)~Projection trajectories of the condensate in the $y$-$z$ plane (red curves). Black arrows indicate the orientation of the synthetic magnetic field, while green arrows show its angular deviation from the $-y$ axis for $\theta = 30^\circ$, $45^\circ$, and $60^\circ$ (from left to right). System parameters are consistent with Fig.~\ref{fig5}. Initial perturbations differ between methods: GP simulations implement a spatial displacement $ y_0 = 3k_0^{-1} $ along the $y$-axis, whereas the HD theory uses the initial condition $ \epsilon_2 = 3\left(\frac{4}{15}\pi R_x R_y^2 R_z\right)^{-1} $ as given in Eq.~\eqref{ansatzdipole}.
        }
    \label{fig6}
\end{figure*}
%%%%%%%%%%%%%%%%%%%%%%%%%%%%%%

To investigate the dipole dynamics of the condensate, which characterize the center-of-mass oscillations after being subjected to external perturbations, we analyze the system by introducing small perturbations in the density $ \delta n(\mathbf{r},t) $ and phase $ \delta\phi(\mathbf{r},t) $. These perturbations take the form:  
\begin{equation}
\label{ansatzdipole}
\begin{aligned}
    \delta n &= \epsilon_1\frac{x}{R_x} + \epsilon_2\frac{y}{R_y} + \epsilon_3\frac{z}{R_z},\\
    \quad \delta\phi &= \frac{g}{\hbar}\left(\alpha_1\frac{x}{R_x} + \alpha_2\frac{y}{R_y} + \alpha_3\frac{z}{R_z}\right),
\end{aligned}
\end{equation}  
where the coefficients $\epsilon_i$ and $\alpha_i$ ($i=1,2,3$) depend on time. Using the expression for the total density $n(x,y,z,t) = n_0 + \delta n$, together with the Thomas-Fermi profile for the equilibrium density $n_0$ and the perturbation form $\delta n$ given in Eq.~\eqref{ansatzdipole}, it is straightforward to calculate:
\begin{equation}
\label{dipoleexpect}
\begin{aligned}
\langle x\rangle &= \frac{4}{15}\pi R_x^2 R_y R_z \epsilon_1, \\
\langle y\rangle &= \frac{4}{15}\pi R_x R_y^2 R_z \epsilon_2, \\
\langle z\rangle &= \frac{4}{15}\pi R_x R_y R_z^2 \epsilon_3.
\end{aligned}
\end{equation}
Therefore, the dipole motion of the system can be characterized by studying the dynamics of the coefficients $\epsilon_1$, $\epsilon_2$, and $\epsilon_3$. To obtain the dynamics for $\epsilon_i$, we substitute the ansatz Eq.~\eqref{ansatzdipole} into the linearized HD equations~\eqref{perturb}, which yields a set of six coupled dynamical equations:
\begin{equation}
\label{dipoleeq}
\begin{aligned}
\frac{d \epsilon_{1}}{d t} &= \frac{m}{m^*}\omega_x^2\alpha_1 -\omega_{\eta_1}\epsilon_2 - \omega_{\eta_2}\epsilon_3, \\
\frac{d \epsilon_{2}}{d t} &= \omega_y^2\alpha_2+ \omega_{\eta_1}\epsilon_1, \\
\frac{d \epsilon_{3}}{d t} &= \omega_z^2\alpha_3+\omega_{\eta_2}\epsilon_1, \\
\frac{d \alpha_{1}}{d t} &=  - \epsilon_1 -\omega_{\eta_1}\alpha_2 - \omega_{\eta_2}\alpha_3, \\
\frac{d \alpha_{2}}{d t} &= - \epsilon_2 + \omega_{\eta_1}\alpha_1, \\
\frac{d \alpha_{3}}{d t} &= - \epsilon_3 + \omega_{\eta_2}\alpha_1.
\end{aligned}
\end{equation}  
Here, $\omega_{\eta_1} \equiv (\omega_{x}/\omega_{y})\omega_{\text{eff,1}}=(\omega_{y}/\omega_{x})\omega_{\text{eff,1}}^\prime$ and $\omega_{\eta_2} \equiv (\omega_{x}/\omega_{z})\omega_{\text{eff,2}}=(\omega_{z}/\omega_{x})\omega_{\text{eff,2}}^\prime$ characterize the coupling coefficients induced by the synthetic magnetic field. According to Eqs.~\eqref{dipoleeq}, the dipole mode along the $x$-direction is simultaneously coupled to both the $y$- and $z$-dipole modes, whereas the $y$- and $z$-dipole modes is coupled to the $x$-dipole mode only. This asymmetric coupling originates from the fact that the spin-orbit coupling is aligned along the $x$-axis, and that the synthetic magnetic field is oriented perpendicular to the $x$-axis.

Under the degeneracy condition, where $\omega_\text{D} \equiv \sqrt{m/m^*}\omega_{x} = \omega_{y} = \omega_{z}$~\cite{PhysRevA.108.053316,PhysRevResearch.7.013219}, two super-modes confined in the $y$-$z$ plane can be constructed by linearly combining the coefficients $\epsilon_2$, $\epsilon_3$, $\alpha_2$, and $\alpha_3$. The corresponding expressions are given by:
\begin{subequations}
\begin{align}
    \epsilon_\perp &= \frac{\omega_{\eta_1}\epsilon_2 + \omega_{\eta_2}\epsilon_3}{\omega_\eta}, \quad \alpha_\perp = \frac{\omega_{\eta_1}\alpha_2 + \omega_{\eta_2}\alpha_3}{\omega_\eta},\label{paralmode1}\\
    \epsilon_\parallel &= \frac{-\omega_{\eta_2}\epsilon_2 + \omega_{\eta_1}\epsilon_3}{\omega_\eta}, \quad \alpha_\parallel = \frac{-\omega_{\eta_2}\alpha_2 + \omega_{\eta_1}\alpha_3}{\omega_\eta},\label{perbmode1}
\end{align}
\end{subequations}
where $\omega_\eta = \sqrt{\omega_{\eta_1}^2 + \omega_{\eta_2}^2}$. 

Here, $\epsilon_\parallel$ corresponds to the longitudinal mode $r_\parallel$, where the subscript $\parallel$ indicates that this mode is oriented parallel to the synthetic magnetic field $\bm{B}$. In contrast, $\epsilon_\perp$ characterizes the transverse mode $r_\perp$, which is perpendicular to $\bm{B}$ and aligned with the detuning gradient $\bm{\nabla}\delta$. In the same time, using the relation $\omega_{\eta_1}/\omega_{\eta} = \sin\theta$, $\omega_{\eta_2}/\omega_{\eta} = \cos\theta$ and the expressions in Eqs.~\eqref{dipoleexpect}, one can find that the two super-modes are expressed by
\begin{equation}
\label{newsupermode}
\begin{aligned}
        \langle r_\perp\rangle &= \langle y\rangle\sin\theta + \langle z\rangle\cos\theta,\\
        \langle r_\parallel\rangle &= -\langle y\rangle\cos\theta + \langle z\rangle\sin\theta.
\end{aligned}
\end{equation}

% Notably, the transverse coefficient $\epsilon_\perp$ couples to $\epsilon_1$, whereas the longitudinal coefficient $\epsilon_\parallel$ remains decoupled from both $\epsilon_1$ and $\epsilon_\perp$.
The dynamics naturally decouple into two distinct situations. One is the transverse dynamics, which involves coupling between $\epsilon_\perp$ and $\epsilon_1$, capturing the interplay between the synthetic magnetic field and spin-orbit coupling. It's governed by the following set of coupled differential equations:
\begin{equation}
\label{perbmode2}
\begin{aligned}
\frac{d^2 \epsilon_{1}}{d t^2}+\omega_\text{D}^2\left(1-\frac{\omega_{\eta}^{2}}{\omega_\text{D}^2}\right)\epsilon_{1}+2\omega_{\eta}\frac{d \epsilon_{\perp}}{d t}	&=0,\\
\frac{d^2 \epsilon_{\perp}}{d t^2}+\omega_\text{D}^2\left(1-\frac{\omega_{\eta}^{2}}{\omega_\text{D}^2}\right)\epsilon_{\perp}-2\omega_{\eta}\frac{d \epsilon_{1}}{d t}	&=0,
\end{aligned}
\end{equation}
which support an elliptical precessional motion within the velocity field plane, reminiscent of a Foucault pendulum.

In contrast, the longitudinal mode $\epsilon_\parallel$—which is aligned with the synthetic magnetic field $\bm{B}$—decouples entirely and exhibits independent harmonic motion. Its dynamics are described by a simple second-order differential equation:
\begin{equation}
\label{paralmode2}
\frac{d^2\epsilon_\parallel}{dt^2} + \omega_\text{D}^2\epsilon_\parallel = 0,
\end{equation}
indicating a standard dipole oscillation at frequency $\omega_\text{D}$, unaffected by the spin-orbit coupling or detuning gradient.

Motivated by the decoupling of transverse and longitudinal super-modes under the degeneracy condition, we design two excitation protocols to probe distinct dynamics. As illustrated in Fig.~\ref{fig4}\hyperref[fig4]{(a)}, the first protocol displaces the condensate along the $x$-axis, selectively exciting the transverse super-mode; the second protocol applies a displacement along the $y$-axis, thereby activating both transverse and longitudinal modes. These protocols enable a direct comparison between coupled precessional motion and decoupled harmonic oscillations, providing clear evidence of the synthetic magnetic field pointing to an arbitrary orientation.

The dynamical response after the $x$-dipole excitation is shown in Fig.~\ref{fig5}. For different values of the orientation parameter $\theta$, the dipole motion exhibits qualitatively similar beating behavior, with only quantitative differences. In particular, the frequency and amplitude of the dipole mode along the $x$-direction remain unchanged with $\theta$. For the $y$- and $z$-dipole modes, analysis based on Eq.~\eqref{newsupermode} shows that the longitudinal mode $r_\parallel$ is not excited. As a result, the condensate undergoes precessional motion confined to the velocity field plane, representing the dynamics of $x$-dipole mode and transverse super-mode $r_\perp$. This motion resembles a Foucault pendulum~\cite{PhysRevLett.120.183202}, characterized by periodic energy exchange between the $y$- and $z$-dipole modes, as determined by the superposition structure in Eq.~\eqref{newsupermode}. Furthermore, we have also performed the GP simulations, which confirm these features and show an excellent agreement with the predictions of HD theory.

Figure~\ref{fig6} displays the system response following $y$-dipole excitation. Unlike the previous case, this protocol activates both super-modes $r_{\perp}$ and $r_\parallel$, leading to a three-dimensional bi-conical trajectory. The elliptical motion within the velocity field plane is now combined with harmonic oscillations along the orientation of the synthetic magnetic field. This results in a characteristic beating pattern and a bi-conical trajectory whose shape depends on the field orientation. When projected onto the $y$-$z$ plane, the motion traces out a tilted bi-conical section. The cone half-angle (green arrows), measured relative to the orientation of the synthetic magnetic field (black arrows), reflects the direction of the detuning gradient $\bm{\nabla}\delta$. Notably, the conical motion preserves azimuthal phase locking, providing a robust and measurable signature of the field orientation. These features are well captured by HD theory and show excellent agreement with GP simulations (see the green dotted line in Fig.~\ref{fig6}\hyperref[fig6]{(c1)}).

\subsection{Quadrupole Oscillation Following a Quench of the Orientation of the Synthetic Magnetic Field}

The investigation of the quadrupole excitation modes provide critical insights of the superfluidity of BEC. These modes can be classified into diagonal quadrupole modes, which are primarily associated with the quadratic operators $x^2$, $y^2$, and $z^2$, and scissors modes, which involve the cross terms $xy$, $xz$, and $yz$. The diagonal quadrupole modes correspond to oscillations in the overall shape of the condensate, whereas the scissors modes correspond to angular rotations of the condensate within a plane~\cite{PhysRevLett.83.4452,PhysRevLett.84.2056,PhysRevA.69.043621}. Recently, in SOC BECs, quadrupole modes have been employed to experimentally observe the Hall effect~\cite{doi:10.1073/pnas.1202579109}. The introduction of a synthetic magnetic field generates a rigid-body-like rotational velocity field, which induces novel coupling mechanisms between the dynamics of different quadrupole modes. This section focuses on the rich nonequilibrium dynamics after abruptly altering the orientation $\theta$ of the synthetic magnetic field.

\textcolor{black}{To investigate the quadrupole dynamics, which describe internal excitations of the condensate, we analyze the system by introducing small perturbations in the density $ \delta n(\mathbf{r},t) $ and phase $ \delta\phi(\mathbf{r},t) $ as follows:}
\begin{equation}
    \begin{aligned}
        \delta n=&\epsilon_1\frac{xy}{R_xR_y}+\epsilon_2\frac{x^2}{R_x^2}+\epsilon_3\frac{y^2}{R_y^2}+\epsilon_4\frac{z^2}{R_z^2}\\
        &+\epsilon_{5}\frac{xz}{R_{x}R_{z}
        }+\epsilon_{6}\frac{yz}{R_{y}R_{z}},\\
        \delta \phi=&\frac{g}{\hbar}\left(\alpha_1\frac{xy}{R_xR_y}+\alpha_2\frac{x^2}{R_x^2}+\alpha_3\frac{y^2}{R_y^2}+\alpha_4\frac{z^2}{R_z^2}\right.\\
        &\left.+\alpha_{5}\frac{xz}{R_{x}R_{z}}+\alpha_{6}\frac{yz}{R_{y}R_{z}}\right).
    \end{aligned}
    \label{raodongnishe}
\end{equation}

Similar to the analysis of dipole modes, the twelve coefficients $\epsilon_i$ and $\alpha_i$ ($i=1,\dots,6$) depend on time. Through the HD theory, the evolution of the condensate’s overall shape and angular rotation can be described by the expectation values of the corresponding quadrupole modes, expressed as follows:
% \begin{equation}
%     \begin{aligned}
%         \langle xy\rangle&=\frac{4\pi}{105}R_x^2R_y^2R_z\epsilon_1,\\
%         \langle x^2\rangle&=\frac{N}{7}R_x^2+\frac{4\pi}{105}R_x^3R_yR_z\left(3\epsilon_2+\epsilon_3+\epsilon_4\right),\\
%         \langle y^2\rangle&=\frac{N}{7}R_y^2+\frac{4\pi}{105}R_xR_y^3R_z\left(\epsilon_2+3\epsilon_3+\epsilon_4\right),\\
%         \langle z^2\rangle&=\frac{N}{7}R_z^2+\frac{4\pi}{105}R_xR_yR_z^3\left(\epsilon_2+\epsilon_3+3\epsilon_4\right),\\
%         \langle xz\rangle&=\frac{4\pi}{105}R_x^2R_yR_z^2\epsilon_5,\\
%         \langle yz\rangle&=\frac{4\pi}{105}R_xR_y^2R_z^2\epsilon_6.
%     \end{aligned}
% \end{equation}
\begin{equation}
    \begin{aligned}
        \langle xy\rangle&=\frac{4\pi}{105}R_x^2R_y^2R_z\epsilon_1,\\
        \langle xz\rangle&=\frac{4\pi}{105}R_x^2R_yR_z^2\epsilon_5,\\
        \langle yz\rangle&=\frac{4\pi}{105}R_xR_y^2R_z^2\epsilon_6,
    \end{aligned}
\end{equation}
and
\begin{equation}
    \begin{aligned}
        \langle x^2\rangle&=\frac{N}{7}R_x^2+\frac{4\pi}{105}R_x^3R_yR_z\left(3\epsilon_2+\epsilon_3+\epsilon_4\right),\\
        \langle y^2\rangle&=\frac{N}{7}R_y^2+\frac{4\pi}{105}R_xR_y^3R_z\left(\epsilon_2+3\epsilon_3+\epsilon_4\right),\\
        \langle z^2\rangle&=\frac{N}{7}R_z^2+\frac{4\pi}{105}R_xR_yR_z^3\left(\epsilon_2+\epsilon_3+3\epsilon_4\right).
    \end{aligned}
\end{equation}
This means that the collective dynamics related to the condensate’s shape deformation and rotational response can be characterized by analyzing the time evolution of these twelve quadrupole coefficients $\epsilon_i$ and $\alpha_i$. Substituting the ansatz~\eqref{raodongnishe} into Eqs.~\eqref{perturb} yields a set of twelve coupled dynamical equations
\begin{widetext}
    \begin{equation}
        \label{quadropole}
        \begin{matrix}
        \begin{aligned}
            \frac{d \epsilon_{1}}{d t}&=\omega_{x y}^{2} \alpha_{1}+2 \omega_{\eta_1} \epsilon_{2}-2 \omega_{\eta_1} \epsilon_{3}{-\omega_{\eta_2}\epsilon_6}, \\
            \frac{d \epsilon_{2}}{d t}&=3 \frac{m}{m^{*}} \omega_{x}^{2} \alpha_{2}+\omega_{y}^{2} \alpha_{3}+\omega_{z}^{2} \alpha_{4}-\omega_{\eta_1}\epsilon_{1}{-\omega_{\eta_2}\epsilon_5}, \\
            \frac{d \epsilon_{3}}{d t}&=\frac{m}{m^{*}} \omega_{x}^{2} \alpha_{2}+3 \omega_{y}^{2} \alpha_{3}+\omega_{z}^{2} \alpha_{4}+\omega_{\eta} \epsilon_{1}, \\
            \frac{d \epsilon_{4}}{d t}&=\frac{m}{m^{*}} \omega_{x}^{2} \alpha_{2}+\omega_{y}^{2} \alpha_{3}+3 \omega_{z}^{2} \alpha_{4}{+\omega_{\eta_2}\epsilon_5}, \\
            \frac{d\epsilon_{5}}{d t}&=\omega_{xz}^{2}\alpha_{5}-\omega_{\eta_1}\epsilon_{6}{+2\omega_{\eta_2}\epsilon_2-2\omega_{\eta_2}\epsilon_4},\\
            \frac{d\epsilon_{6}}{d t}&=\omega_{yz}^{2}\alpha_{6}+\omega_{\eta_1}\epsilon_{5}{+\omega_{\eta_2}\epsilon_1},\\
        \end{aligned}\  &\  \begin{aligned}
            \frac{d \alpha_{1}}{d t}&=2 \omega_{\eta_1} \alpha_{2}-2 \omega_{\eta_1} \alpha_{3}{-\omega_{\eta_2}\alpha_6}-\epsilon_{1}, \\
            \frac{d \alpha_{2}}{d t}&=-\omega_{\eta_1} \alpha_{1}{-\omega_{\eta_2}\alpha_5}-\epsilon_{2}, \\
            \frac{d \alpha_{3}}{d t}&=-\epsilon_{3}+\omega_{\eta_1} \alpha_{1}, \\
            \frac{d \alpha_{4}}{d t}&=-\epsilon_{4}+{\omega_{\eta_2}\alpha_5},\\
            \frac{d\alpha_{5}}{d t}&=-\omega_{\eta_1}\alpha_{6}{+\omega_{\eta_2}\alpha_2-\omega_{\eta_2}\alpha_4}-\epsilon_{5},\\
            \frac{d\alpha_{6}}{d t}&=\omega_{\eta_1}\alpha_{5}{+\omega_{\eta_2}\alpha_1}-\epsilon_{6}.\\
        \end{aligned} \\
        \end{matrix}
    \end{equation}
\end{widetext}

%%%%%%%%%%%%%%%%%%%%%%%%%%%%%%
\begin{figure}[t]
    \centerline{
    \includegraphics[width=0.5\textwidth]{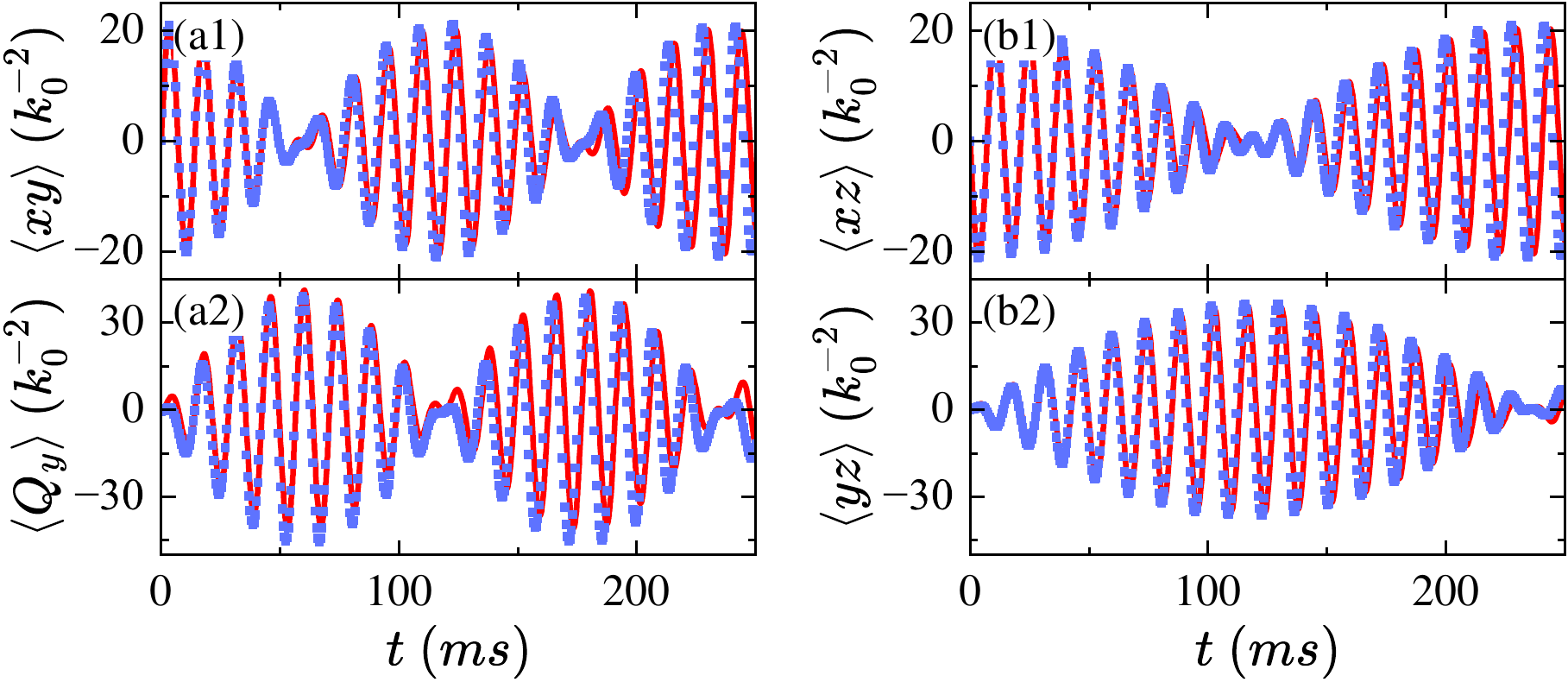}}
        \caption{Collective quadrupole dynamics induced by a quench of the orientation of the synthetic magnetic field from $\theta = 0^\circ$ to $\theta' = 90^\circ$. Panels (a1)-(b2) display the resulting beating dynamics of: (a1) the scissors mode $xy$, (a2) the quadrupole super-mode $Q_y$, (b1) the scissors mode $xz$, and (b2) the scissors mode $yz$. Red curves denote predictions from HD theory, while blue dots represent results from GP simulations, showing good agreement. The system parameters are: trapping frequencies $(\omega_x, \omega_y, \omega_z) = 2\pi \times (50\sqrt{3}, 50, 50)$ Hz, detuning gradient $\eta = 0.001 E_r$, Raman coupling strength $\Omega = 6 E_r$, and particle number $N = 5 \times 10^4$. Initial perturbations in the HD approach are set using the ansatz $\delta\phi$ in Eqs.~\eqref{raodongnishe}, with $\alpha_1(0) = -\alpha_y(\theta = 90^\circ) R_x R_y \hbar / g$ and $\alpha_5(0) = \alpha_z(\theta = 0^\circ) R_x R_z \hbar / g$.}
    \label{fig7}
    \end{figure}
%%%%%%%%%%%%%%%%%%%%%%%%%%%%%%

Previous studies~\cite{PhysRevA.108.053316,PhysRevResearch.7.013219} have revealed that at the specific orientation $\theta = 90^\circ$, the scissors mode $xy$ exhibits strong coupling with a quadrupole \textcolor{black}{super-mode} defined as \textcolor{black}{
\begin{equation}
    Q_y \equiv \frac{\omega_x}{\omega_y}x^2 - \frac{\omega_y}{\omega_x}y^2.
\end{equation}
This coupling gives rise to a perfect beating effect between the scissors mode $xy$ and the super-mode $Q_y$.} Meanwhile, the remaining scissors modes $xz$ and $yz$ become coupled, and their spatial dynamics exhibit a gyroscopic effect. For our analysis, the post-quench orientation angle is set to $\theta^\prime = 90^\circ$, in line with previous studies, allowing for direct comparison and intuitive interpretation of the resulting mode couplings.

The quadrupole modes exhibit very different dynamics for different orientation $\theta$ of the synthetic magnetic field, which are fully described by Eqs.~\eqref{quadropole}. To probe quench dynamics, we consider a sudden change of the field orientation from $\theta_0 = 0^\circ$ (along $-y$) to $\theta^\prime = 90^\circ$ (along $+z$), as shown in Fig.~\ref{fig4}\hyperref[fig4]{(b)}. This abrupt change drives the system out of equilibrium and initiates collective oscillations. The condensate is initially prepared in the equilibrium state corresponding to $\theta_0$, with a phase profile $\phi= \alpha_z(\theta = 0^\circ) xz$. After the quench, since the system cannot immediately adapt to the new equilibrium at $ \theta^\prime $, the deviation defines the initial set of the coefficients:
\begin{equation}
    \begin{aligned}
        \alpha_1(t=0) &= -\alpha_{y}(\theta = 90^\circ) R_xR_y\frac{\hbar}{g}, \\
        \alpha_5(t=0) &= \alpha_{z}(\theta = 0^\circ) R_xR_z\frac{\hbar}{g},
    \end{aligned}
\end{equation}
with all other $\alpha_i$ and $\epsilon_i$ initialized to zero at $t = 0 $. The entire process resembles a classical system composed of a ball and a spring, where the spring constant is suddenly changed at equilibrium, causing the system to enter a dynamical evolution.

Figure~\ref{fig7} demonstrates the evolution of quadrupole modes following the quench. As shown in Fig.~\ref{fig7}\hyperref[fig7]{(a1)-(a2)}, a perfect beating effect emerges between the $xy$ scissors mode and the quadrupole super-mode $Q_y$. A similar response is observed for the $xz$ and $yz$ scissors modes in Fig.~\ref{fig7}\hyperref[fig7]{(b1)-(b2)}. When all four modes are simultaneously excited, the condensate exhibits dynamics that closely follow the predictions of HD theory and GP simulations during the first 100 ms, with gradual deviations appearing at later times. In real space, the condensate alternates periodically between expansion-contraction and rotational motion in the $x$-$y$ plane, driven by a coupling strength of $2\omega_\eta$, while also undergoing gyroscopic precession about the $z$-axis with the coupling strength $\omega_\eta$. These distinct coupling strengths manifest in the beating period: the evolution shown in Fig.~\ref{fig7}\hyperref[fig7]{(a)} has twice the frequency of that in Fig.~\ref{fig7}\hyperref[fig7]{(b)}. The distinct behaviors observed are direct consequences of the geometric reconfiguration imposed by the sudden switch to the $\theta^\prime = 90^\circ$ field orientation. These results demonstrate that the orientation of the synthetic magnetic field serves as a control parameter that dictates the pattern of mode coupling and the resulting dynamics in non-equilibrium quadrupole oscillations. This highlights the crucial role of field geometry in shaping the collective behavior of SOC BECs.

\section{Conclusions}\label{sec5}

Within the spinor HD theory, we systematically investigate both the ground-state configuration and the collective oscillations of BECs in a position dependent detuning. The position-dependent detuning can induces a synthetic magnetic field with tunable orientation, which is always perpendicular to the detuning gradient. In the ground state, the condensate exhibits a rigid-body-like velocity field. The rotational behavior is characterized by two key quantities: the angular velocity and the angular momentum. For general orientation parameters ($\theta \ne 0^\circ, 90^\circ$), trap anisotropy causes angular velocity and angular momentum to deviate from the orientation of the synthetic magnetic field by two distinct finite angles. These three vectors become aligned under axially symmetric condition (e.g., $\omega_y = \omega_z$) or at specific orientation parameters ($\theta = 0^\circ$, $90^\circ$). 

Under the degeneracy condition $\omega_\text{D} \equiv \sqrt{m/m^*}\omega_x = \omega_y = \omega_z$, which satisfies the axial symmetry, our dynamical investigations reveal that, regardless of the orientation angle, the dipole motion of the condensate—representing its center-of-mass dynamics—can always be decomposed into three distinct modes: (i) a dipole mode along the spin-orbit coupling direction, (ii) a transverse super-mode aligned with the detuning gradient, and (iii) a longitudinal super-mode oriented along the synthetic magnetic field. The first two modes are coupled and give rise to a precessional motion confined within the velocity field plane, while the third remains decoupled, undergoing independent harmonic oscillations. Depending on the excitation protocol, the resulting dynamics can manifest as two-dimensional Foucault-like precession or a three-dimensional bi-conical trajectory. Furthermore, by quenching the orientation of the synthetic magnetic field, different quadrupole modes are excited simultaneously, leading to characteristic spatial deformation-rotation dynamics of the condensate. In all cases, the nonequilibrium behavior obtained from GP simulations shows excellent agreement with predictions from HD theory. 

Our findings indicate that SOC BECs may serve as sensitive probes for magnetic field gradients, providing information about both magnitude and orientation. Our work offers valuable insights into the dynamics induced by the synthetic magnetic field and contributes to the theoretical foundation for quantum simulation and sensing applications.

\begin{acknowledgments}
This work is supported by the National Natural Science Foundation of China (NSFC) under Grants No. 112374247 and No. 11974235, as well as by the Shanghai Municipal Science and Technology Major Project (Grant No. 2019SHZDZX01-ZX04). C.Q. is supported by ACC-New Jersey under Contract No. W15QKN-18-D-0040.
\end{acknowledgments}

% \nocite{*}
\bibliography{ref.bib} 

\end{document}